# Exact solutions for discrete breathers in forced – damped chain.

O.V.Gendelman

Faculty of Mechanical Engineering, Technion – Israel Institute of Technology

e-mail: ovgend@tx.technion.ac.il

Exact solutions for symmetric discrete breathers (DBs) are obtained in forced – damped linear chain with on-site vibro-impact constraints. The damping is related to inelastic impacts; the forcing may be chosen from broad class of periodic antisymmetric functions. Global conditions for existence and stability of the DB are established. Some unusual phenomena, like non-monotonous dependence of the stability boundary on the forcing amplitude, are revealed analytically for the full system and illustrated numerically for small periodic lattices.

PACS numbers: 63.20.Rw, 63.20.Ry, 05.45.Yv

Discrete breathers (DBs) or intrinsic localized modes (ILMs) are common for many nonlinear lattices [1]. Generally, they are exponentially localized (if a coupling between the neighbors in the lattice has linear component) and can demonstrate remarkable stability also for two – and three – dimensional applications [2]. Numerous systems which exhibit the DBs include chains of mechanical oscillators [3], superconducting Josephson junctions [4], nonlinear magnetic metamaterials [5], electrical lattices [6], michromechanical cantilever arrays [7], antiferromagnets [8] and many other physical systems [9].

The DBs in Hamiltonian models of the discrete nonlinear lattices have received a lot of attention and powerful methods were devised for analysis of their structure and stability. Still, in many applications the damping cannot be neglected; in order to maintain the DB, one should compensate it by some kind of direct or parametric external forcing [2]. Existence of such DBs has been proven in "anticontinuum" limit [10] and their properties were extensively studied by numeric continuation of the solution from the limit of uncoupled oscillators [11].

Lack of Hamiltonian structure changes the properties of the DB. Instead of continuous family of localized solutions, one expects to obtain a discrete set of attractors. Consequently, many of methods devised for computation and analysis of the Hamiltonian DBs are not applicable in the forced- damped systems. In particular, there exist some models where the solutions for discrete Hamiltonian DBs can be computed exactly [12, 13]. To the best of the author's knowledge, no such exact solutions are known in the forced – damped discrete lattices. Exact solutions for forced - damped breathers were obtained only for some continuous models [14]. This Letter is devoted exactly to this problem and suggests a model which allows derivation of exact solutions for the forced – damped DBs and study of some of their global properties.

This model is a homogeneous chain with linear nearest – neighbor interactions; in addition, each particle can move only between inelastic impact constraints. Different models based on vibro – impact lattices and hard potentials were widely used in a context of the discrete breathers and other related problems in solid – state physics [13, 15], but not for the problem of forced-damped DB. Sketch of the system is presented in Fig. 1

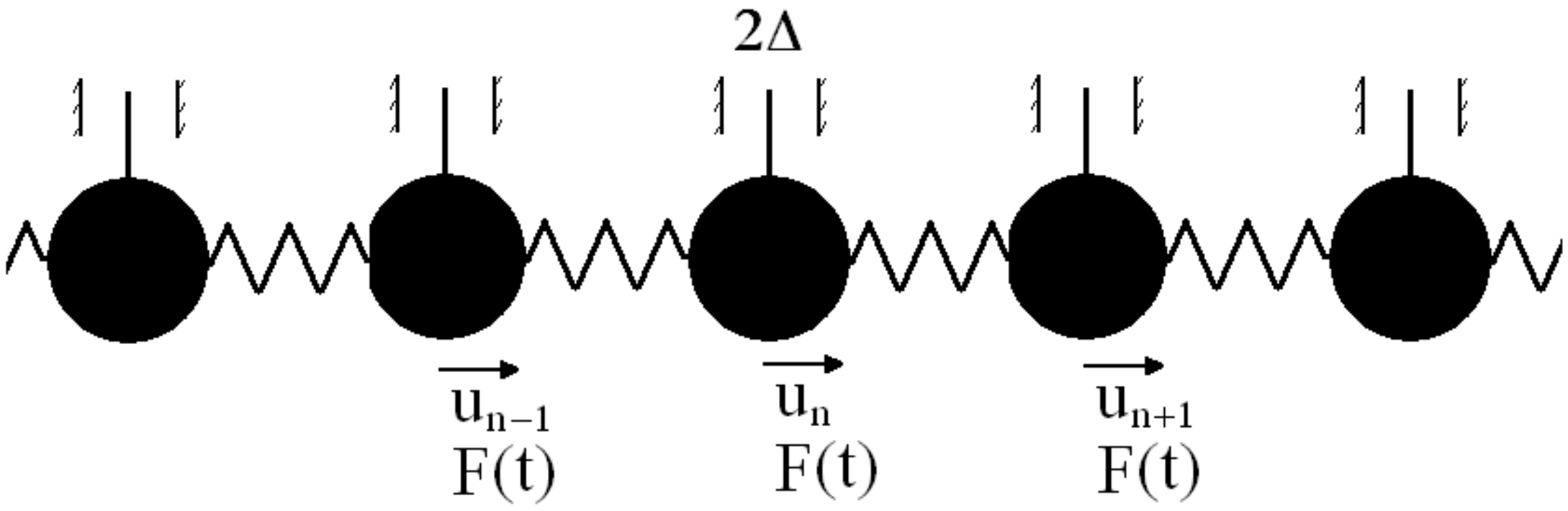

Figure 1. Sketch of the system.

Each particle is subject to the same external forcing. At this stage, the forcing function is considered to be periodic and antisymmetric. Between impacts, displacement $u_n$ of the $n^{th}$ particle is described as:

$$\ddot{u}_n + \gamma(2u_n - u_{n-1} - u_{n+1}) = F(t),\ \ F(t) = F(t+2\pi),\ F(t+\pi) = -F(t), t \in \mathbb{R} \tag{1}$$

Without affecting the generality, the clearance Δ and the particle mass are set to unity, period of the external forcing is equal to 2π, and the only parameter which characterizes the chain is a coupling coefficient γ. Equations (1) are valid between successive impacts, i.e. as $|u_n| < 1$. If the particle n impacts the constraint at time instance $t_{i,n}$, the following conditions are satisfied:

$$t = t_{i,n};\ u_n(t_{i,n}) = \pm 1;\ \dot{u}_n(t_{i,n} + 0) = -k\dot{u}_n(t_{i,n} - 0),\ \ 0 \le k \le 1 \tag{2}$$

Here k is a restitution coefficient. Notation ±0 is used for the particle velocities immediately after (+) and before (-) the impact. Values of k<1 correspond to inelastic impacts and it is the only source of damping in the model.

The model is obviously non-integrable and not solvable in a general form. Therefore, first the topology of the DB is defined and then the solution is constructed. Here we consider the simplest one – site symmetric DB. We adopt that responses of all particles are periodic and symmetric, and only particle with n=0 impacts each of its constraints symmetrically with period 2π:

$$\begin{aligned} &u_n(t)=u_n(t+2\pi m)=-u_n(t+\pi(2m+1)), n,m\in\mathbb{Z}, t\in\mathbb{R};\ t_{i,0}=\varphi+\pi m \\ &u_0(\varphi)=1, \dot{u}_0(\varphi+0)=-k\dot{u}_0(\varphi-0);\ u_0(\varphi+\pi)=-1, \dot{u}_0(\varphi+\pi-0)=-\dot{u}_0(\varphi-0) \end{aligned} \tag{3}$$

Here φ is a phase of the impacts. Effect of each impact is equivalent to transfer of certain amount of momentum to the impacting particle. Consequently, the solution we look for should satisfy the following modification of Equation (1) (in terms of distributions):

$$\begin{aligned} &\ddot{u}_n+\gamma(2u_n-u_{n-1}-u_{n+1})=F(t)+2p\delta_{n0}\sum_{j=-\infty}^{\infty}[\delta(t-\varphi+\pi(2j+1))-\delta(t-\varphi+2\pi j)] \\ &|u_n|<1, n\in\mathbb{Z}/0 \end{aligned} \tag{4}$$

δ(t) is Dirac delta – function, $\delta_{nm}$ is Kronecker symbol. The second condition in (4) reflects the fact that the particles with n≠0 never impact the constraints. 2p is yet unknown absolute amount of momentum transferred to the particle with n=0 in the course of each impact. For further analysis, the variables are changed as follows:

$$u_n=v_n+f(t),\ \ddot{f}(t)=F(t) \tag{5}$$

Due to the antisymmetry of *F(t),* it is always possible to find unique *f(t)* satisfying Eq. (5) and the antisymmetry condition $f(t+\pi)=-f(t)$. In terms of new variables, the external forcing term in the first equation of (4) disappears:

$$\ddot{v}_n+\gamma(2v_n-v_{n-1}-v_{n+1})=2p\delta_{n0}\sum_{j=-\infty}^{\infty}[\delta(t-\varphi+\pi(2j+1))-\delta(t-\varphi+2\pi j)], n\in\mathbb{Z} \tag{6}$$

In equation (6) the only forcing term is related to the impacts. The stationary regime is possible only if the impact is symmetric in terms of variables $v_n(t)$:

$$\dot{v}_0(\varphi-0) = -\dot{v}_0(\varphi+0) = p$$

With account of this relationship, equations (2) of the inelastic impact are rewritten as:

$$v_0(\varphi) + f(\varphi) = 1;\ \dot{u}_0(\varphi+0) = -p + \dot{f}(\varphi) = -k(p + \dot{f}(\varphi)) \Rightarrow \dot{f}(\varphi) = p\frac{1-k}{1+k} \tag{7}$$

Equations (7) have 3 unknowns – p, φ and $v_0(\varphi)$. To close the system, one needs the relationship between p, φ and the value $v_0$ in time instance of the impact $v_0(\varphi)$.The latter is obtained by rewriting (6) with the help of formal Fourier series:

$$\ddot{v}_n + \gamma(2v_n - v_{n-1} - v_{n+1}) = -\frac{4p\delta_{n0}}{\pi}\sum_{l=0}^{\infty}\cos((2l+1)(t-\varphi)) \tag{8}$$

It is easy to derive the steady – state solution of (8) in a form:

$$v_n(t) = \frac{4p}{\pi}\left(\frac{-1}{2\gamma}\right)^n \sum_{l=0}^{\infty} \frac{\left((2l+1)^2 - 2\gamma - \sqrt{(2l+1)^4 - 4\gamma(2l+1)^2}\right)^{|n|}}{\sqrt{(2l+1)^4 - 4\gamma(2l+1)^2}} \cos((2l+1)(t-\varphi) \tag{9}$$

Fourier series (9) converges for all n and for $\gamma \in [0, 0.25)$. This interval of the coupling parameter corresponds to *the attenuation zone* of the linear chain – the DB of the considered type can exist only there. In general, some harmonics of the DB in the forced – damped systems can stay in the propagation zone [11], but such solutions are beyond the scope of this work. The expression for $v_0(\varphi)$ is written as:

$$v_0(\varphi) = p\chi(\gamma),\ \chi(\gamma) = \frac{4}{\pi}\sum_{l=0}^{\infty}\frac{1}{\sqrt{(2l+1)^4 - 4\gamma(2l+1)^2}} \tag{10}$$

With account of (10), System (7) can be closed:

$$p\chi(\gamma)+f(\varphi)=1,\ \ \dot{f}(\varphi)=p\frac{1-k}{1+k} \tag{11}$$

Then, the exact solution for the symmetric one – site DB is expressed as:

$$u_n(t)=f(t)+\frac{4p}{\pi}\left(\frac{-1}{2\gamma}\right)^n\sum_{l=0}^{\infty}\frac{\left((2l+1)^2-2\gamma-\sqrt{(2l+1)^4-4\gamma(2l+1)^2}\right)^{|n|}\cos((2l+1)(t-\varphi))}{\sqrt{(2l+1)^4-4\gamma(2l+1)^2}} \tag{12}$$

Parameters *p* and φ can be determined from Equation (11) for specific choice of *F(t).*

In order to study the DB described by (11-12) in some depth, we look at the case of simple harmonic forcing, which allows explicit solution of Equations (11):

$$\begin{gathered}F(t)=-a\cos t\Rightarrow f(t)=a\cos t;\ \ -a\cos\varphi=p\chi(\gamma)-1,-a\sin\varphi=pq,q=\frac{1-k}{1+k}\\ p=\frac{\chi(\gamma)\pm\sqrt{\chi^2(\gamma)a^2-q^2(1-a^2)}}{\chi^2(\gamma)+q^2},\varphi=-\arcsin\frac{pq}{a}\end{gathered} \tag{13}$$

Here *a* is the amplitude of harmonic forcing. Stable branch corresponds to the positive sign in the expression for p. From (13) one obtains the condition for minimum forcing amplitude to provide the DB solution, as a function of the restitution coefficient and the coupling:

$$a\geq\frac{q}{\sqrt{\chi^2(\gamma)+q^2}} \tag{14}$$

To ensure self – consistency of the solution, the neighboring particle should not reach the constraint. Then, for any time instance one should obtain:

$$|u_1(t)| = \left| a\cos t - \frac{2p}{\gamma\pi}\sum_{l=0}^{\infty}\frac{(2l+1)^2-2\gamma-\sqrt{(2l+1)^4-4\gamma(2l+1)^2}}{\sqrt{(2l+1)^4-4\gamma(2l+1)^2}}\cos((2l+1)(t-\varphi)\right| < 1 \qquad (15)$$

Conditions (14) and (15) determine the zone of existence of the DB at the plane of parameters. This zone will be illustrated below in Fig. 3, together with the results on stability of the DBs.

One expects the DB to be a hyperbolic attractor. In order to study stability of the DB, a monodromy matrix of the DB in a chain with N particles and periodic boundary conditions is computed and then the stability patterns for growing N are investigated. Due to simplicity of the model, the monodromy matrix may be explicitly written as:

$$\mathbf{M} = (\mathbf{LS})^2 \qquad (16)$$

Here matrix **L** describes linear dynamics of the system between the impacts; **S** is "saltation matrix" which takes into account the effect of discontinuities on linear perturbation of the periodic orbit in the state space [16]. For given model, these 2N*2N matrices are expressed as follows:

$$\mathbf{L} = \exp(\pi\mathbf{A});\omega = \ddot{u}_0(\varphi-0)$$

$$\mathbf{A} = \begin{pmatrix} 0_{N\times N} & I_{N\times N} \\ N\left\{\begin{matrix} -2\gamma & \gamma & 0 & \cdots & 0 & \gamma \\ \gamma & -2\gamma & \gamma & \ddots & \cdots & 0 \\ 0 & \gamma & \ddots & \ddots & 0 & \vdots \\ \vdots & 0 & \ddots & \ddots & \gamma & 0 \\ 0 & \vdots & \ddots & \gamma & -2\gamma & \gamma \\ \gamma & 0 & \cdots & 0 & \gamma & -2\gamma \end{matrix}\right. & 0_{N\times N} \end{pmatrix}, \mathbf{S} = \begin{pmatrix} \begin{matrix} -k & 0 & \cdots & 0 \\ 0 & 1 & \ddots & \vdots \\ \vdots & \ddots & \ddots & 0 \\ 0 & \cdots & 0 & 1 \end{matrix} & 0_{N\times N} \\ \begin{matrix} \dfrac{\omega(1+k)}{\dot{u}_0(\varphi-0)} & 0 & \cdots & 0 \\ 0 & 0 & \ddots & \vdots \\ \vdots & \ddots & \ddots & \vdots \\ 0 & \cdots & \cdots & 0 \end{matrix} & \begin{matrix} -k & 0 & \cdots & 0 \\ 0 & 1 & \ddots & \vdots \\ \vdots & \ddots & \ddots & 0 \\ 0 & \cdots & 0 & 1 \end{matrix} \end{pmatrix}$$

Values of velocity and acceleration of the impacting particle in the moment of impact are taken from exact solution (12-13). Two generic scenarios of the loss of stability were revealed. The first one corresponds to the transition of two complex conjugate eigenvalues through the unit circle and is related to Neimark – Sacker bifurcation. This bifurcation results in appearance of a quasiperiodic DB. The other scenario corresponds of passage of single eigenvalue through unity. This scenario corresponds to pitchfork bifurcation; the latter results in appearance of a pair of asymmetric DBs. Both bifurcations may be sub- or supercritical; exploration of this issue requires computation of high – order normal forms [16] and is beyond the scope of current Letter. Typical structure of eigenvalues of the monodromy matrix for both these bifurcation scenarios is presented in Figs. 2 and 3.

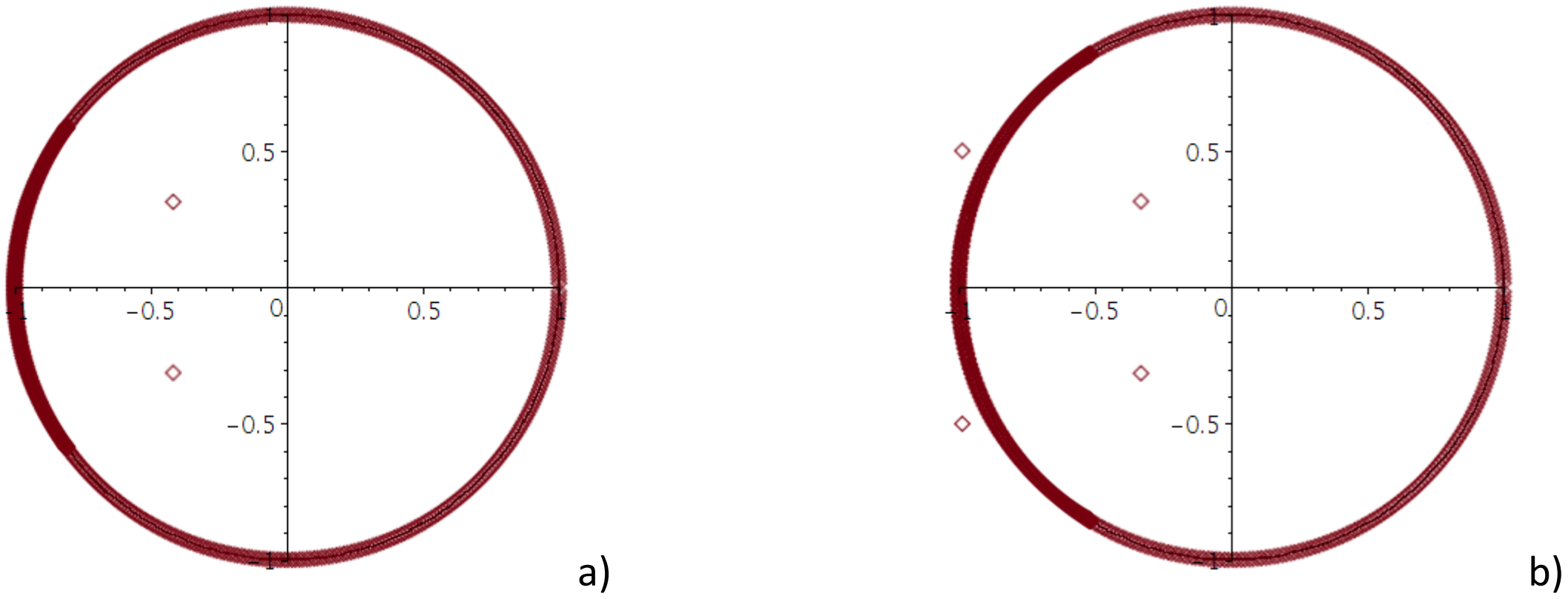

Figure 2. Evolution of eigenvalues of the monodromy matrix corresponding to Neimark – Sacker bifurcation.  N=400, a=0.2,k=0.8,a)γ=0.09, b)γ=0.11.

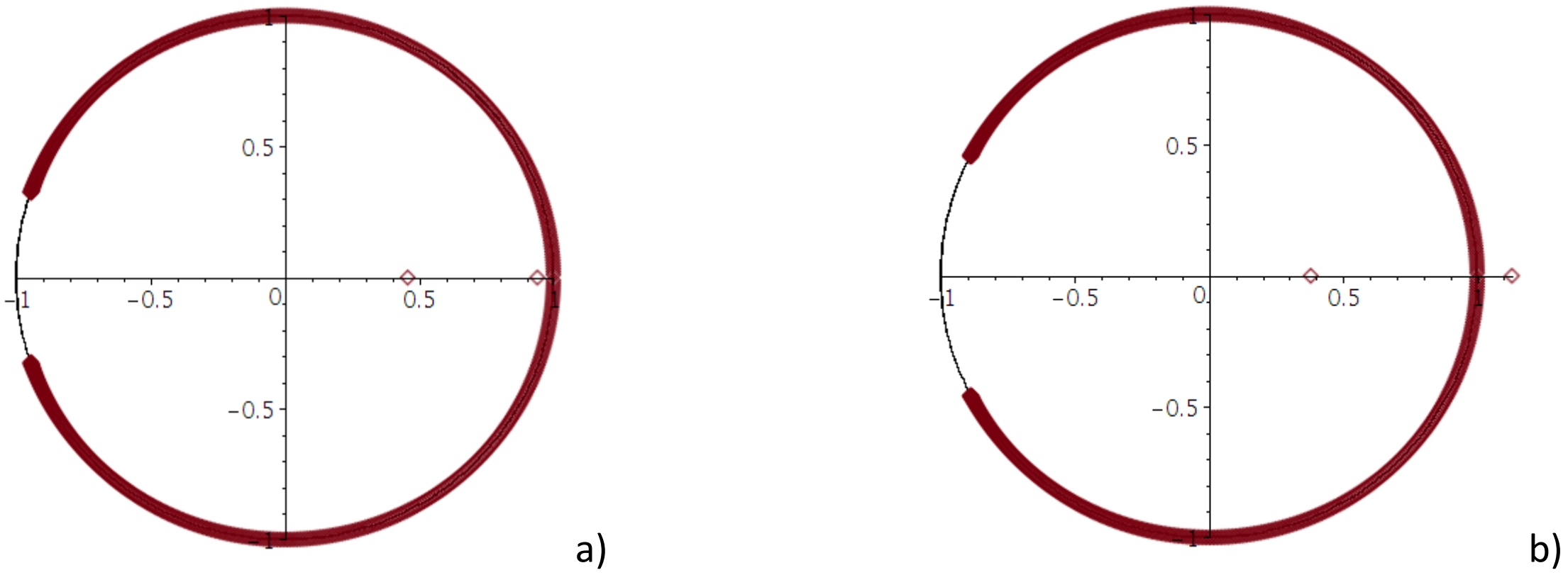

Figure 2. Evolution of eigenvalues of the monodromy matrix corresponding to pitchfork bifurcation. N=400, a=0.8, k=0.8, a)γ=0.05, b)γ=0.045.

For number of particles N>200 the qualitative structure of eigenvalues almost does not depend on N. So, for instance, it is possible to infer from Fig. 2 that for the set of parameters corresponding to the case a) the DB will be stable, and for the case b) – unstable. Zones of existence and stability for the DBs are illustrated in Fig. 4. The lower boundary is described by Eq. (14), the upper one – by Eq. (15).

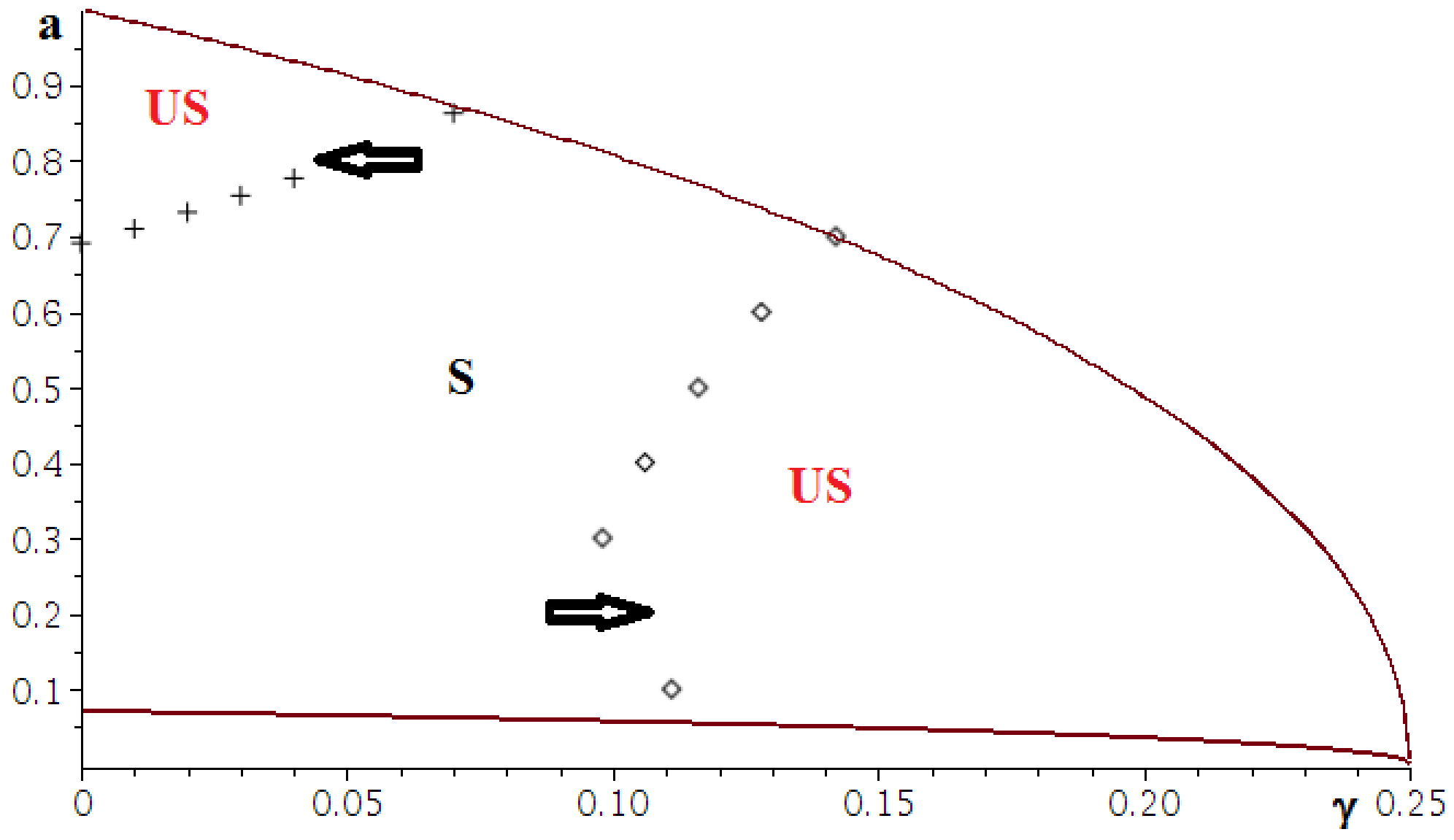

a)

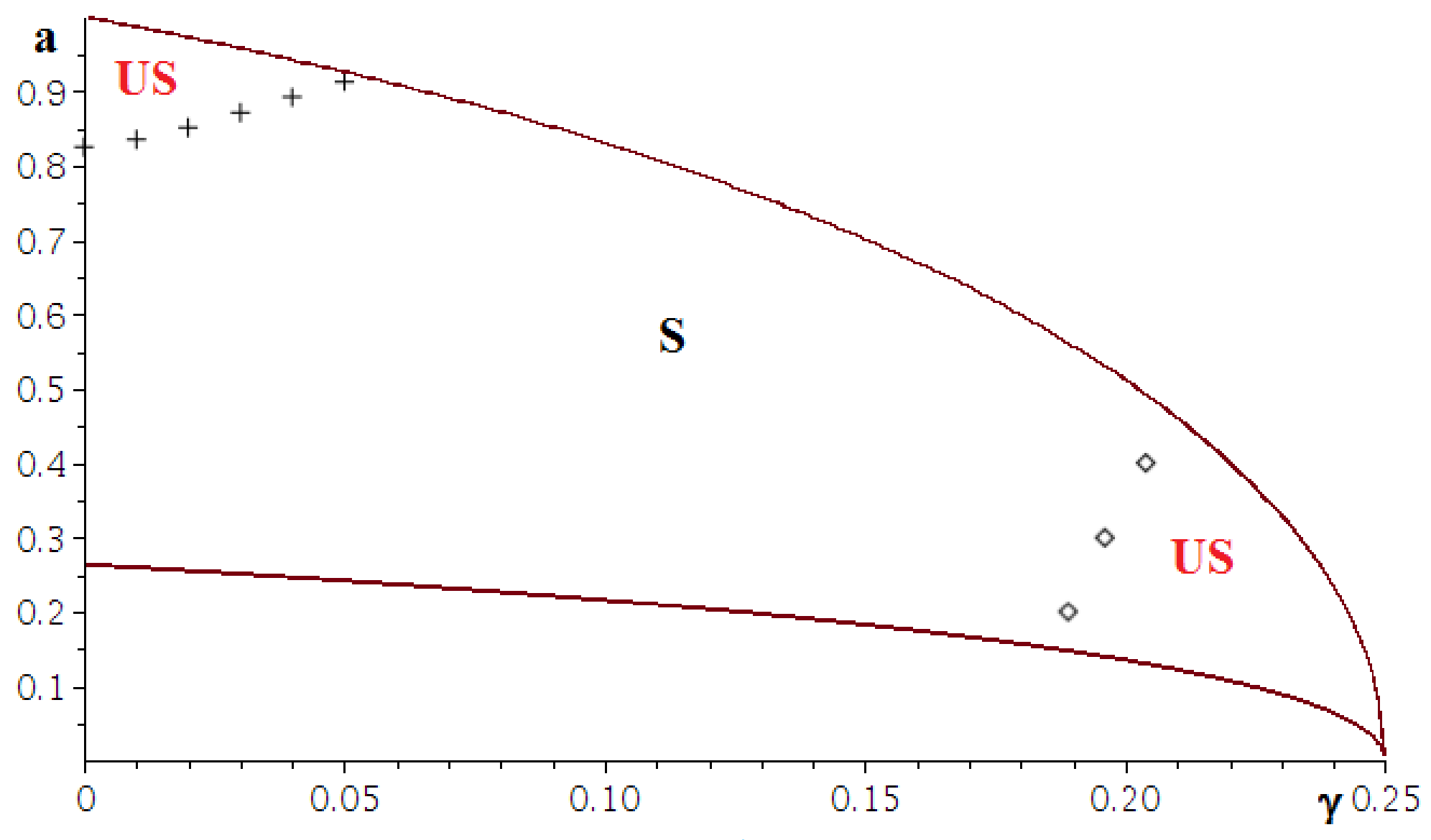

b)

Figure 4. Zones of existence and stability of the symmetric DB on the plane of parameters *γ-a* for two different values of the restitution coefficient, a) k=0.8; b) k=0.4. Crosses denote the pitchfork bifurcation line, diamonds – the line of Neimark – Sacker bifurcation. Transitions corresponding to illustrations in Figs 2 and 3 are marked by arrows in Fig. 4a. “S” and “US” mark zones of stability and instability respectively.

One can see that for lower value of the restitution coefficient, the DB exists for narrower range of the amplitude of external forcing, but is stable for wider range of coupling. It is interesting to mention that for k=0.8 the stability threshold behaves in non-monotonic manner on the plane of parameters.

Normally, experiments with the DBs are performed with relatively small number of particles or nodes. So, in order to illustrate the analytic findings, the numeric simulations were performed for a system with N=20 particles and periodic boundary conditions. Besides, the inelastic impacts were simulated by smooth potential and strongly nonlinear damping [17] and small amount of linear damping was applied at each particle. The system was initialized in accordance with solution (12-13). It was simulated by simple Runge – Kutta method for the restitution coefficient k=0.8, coupling γ=0.104 and three different values of the external forcing –a= 0.1, 0.2, 0.7. From Fig. 4 it is clear that these values of the forcing and coupling are quite close to the boundary of stability. The results of simulation are presented in Fig. 5.

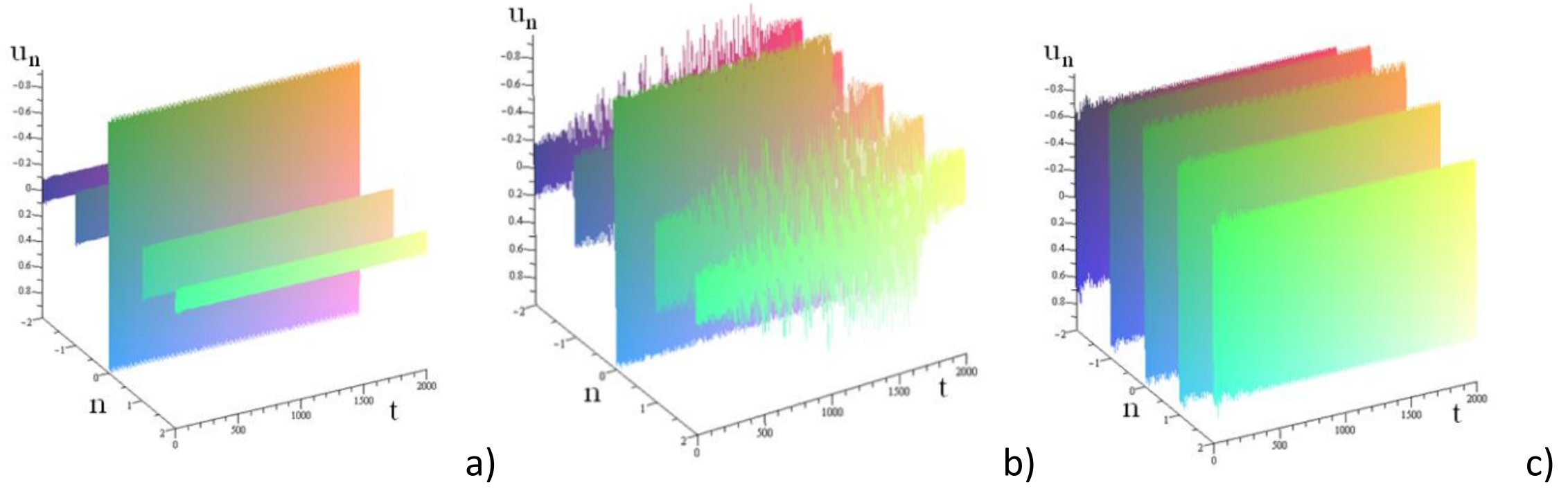

Figure 5. Numeric simulation of the DBs in small system with smoothened potentials and damping, N=20, k=0.8, γ=0.104, a) a=0.1, b) a=0.2, c) a=0.7. Time series for 5 particles (-2≤n≤2) are presented.

For cases a) and c), the system settles in the attractor corresponding to the symmetric DB after some initial transient. In these cases, the time series are very close to solution (12-13). For the case b), the system preserves localization for some time interval, but then the regime becomes delocalized. The results of simulations completely correspond to predictions in Fig. 4a, including the non-monotonous behavior of the stability threshold.

We see that the model allows derivation of exact analytic solutions for discrete breathers in forced – damped essentially nonlinear chain, without any simplifications or approximations. The complete zone of existence and stability of the DB solution in the space of parameters has been derived. It is clear intuitively that that the DB can exist in certain diapason of amplitudes of the external forcing; the results obtained make this statement clear. Two main mechanisms of the loss of stability were revealed. It is interesting to mention that, contrary to some other systems [11] increase of the coupling may play stabilizing role. It is easy to derive from (12-13) that close to the boundary of the propagation zone, where one expects the continuum limit to work, the interval of the forcing amplitudes for which the symmetric DB exists, shrinks as $\Delta a \sim \sqrt{0.25-\gamma}$ . In addition, close to this limit the DB solutions of the investigated type are unstable. Non-monotonic behavior of the stability boundary in the space of parameters was revealed analytically and confirmed numerically for small systems with periodic boundary conditions.

The author is very grateful to Dr. Yuli Starosvetsky for useful discussions and suggestions.

References.